\documentclass[11pt]{article}
\usepackage{amsmath}
\usepackage{graphicx}

\newcommand{\be}{\begin{equation}}
\newcommand{\ee}{\end{equation}}
\newcommand{\bea}{\begin{eqnarray}}
\newcommand{\eea}{\end{eqnarray}}
\newcommand{\ba}{\begin{eqnarray}}
\newcommand{\ea}{\end{eqnarray}}

\newcommand{\beq}{\begin{equation}}
\newcommand{\eeq}{\end{equation}}
\newcommand{\beqa}{\begin{eqnarray}}
\newcommand{\eeqa}{\end{eqnarray}}
\newcommand{\beqar}{\begin{eqnarray*}}
\newcommand{\eeqar}{\end{eqnarray*}}

\def\ci{\cite}
\def\la{\label}
\def\bib{\bibitem}
\def\fr{\frac}
\def\le{\left}
\def\ri{\right}
\def\non{\nonumber }

\def\pu{\partial_\mu}
\def\pU{\partial^\mu}

\def\vp{\varphi}

\def\Om{\Omega_{m}}
\def\Omp{\Omega_\phi}

\def\rvp{\rho_\vp}
\def\wvp{w_\vp}

\def\Omvp{\Omega_\vp}
\def\rp{\rho_\phi}

\def\wpe{w_{\phi eff}}
\def\wvpe{w_{\vp eff}}

\def\N{{\cal N}_4}

\title{Cosmological consequences of scalar mesons from gauge/gravity correspondence}
\author{Axel de la Macorra\footnote{Part of the collaboration Instituto Avanzado de Cosmologia.}  and Leonardo Pati\~no\footnote{Part of the collaboration Instituto Avanzado de Cosmologia.}
\\
Instituto de F\'\i sica,\\ Universidad Nacional Aut\'onoma de M\'exico,\\ Apdo. Postal 20-364, 01000, D.F. M\'exico}

\date{\today}

\begin{document}

\maketitle

\begin{abstract}
We consider the spectrum of mesons for the gauge theory dual to a
supergravity configuration of intersecting D3/D7 branes
\cite{Hovdebo:2005hm}, and use the expression for the Lagrangian of
the scalar mesons to compute explicitly the Lagrangian for the
lightest states in the infrared limit. Assuming that the matter
content of this gauge theory is part of a hidden sector, which interacts
with the standard model only via gravity,   we explore
the cosmological consequences of these lightest scalar mesons for a
FRW universe. We show that phantom fields may appear naturally in
this kind of scenarios.

\end{abstract}

\section{Introduction}

The standard Big-Bang model of cosmology  has been very successful in explaining
the cosmological observations  \ci{WMAP,SN}, however the model needs to be extended
to include at least two periods of positive  acceleration of our universe,
one at high temperatures denoted by Inflation and another at present time
and given in terms of Dark Energy.
Scalar fields are perhaps the best candidates to explain these periods
of acceleration \ci{scalar}. The effect of scalar fields in the cosmological evolution of our
universe has been the object of extensive studies \ci{scalar,Q.ax}.
The equation of state of Dark Energy $w=p/\rho$ is very close to -1
but the cosmological data  \ci{WMAP,SN} seems to favor a $w$   smaller than -1 \ci{DE}.
This can be achieved through the use of the so called phantom fields, which are
scalar fields with a negative kinetic term \ci{phantom}, or alternatively
by the inclusion of an interaction term
between the Dark Energy  scalar field and some other particle \ci{wapp,IDE,IDE.ax}.
Despite their success, these
models have to provide a very particular potential for the scalar
fields, and in most of the cases these potentials are not deduced
from considerations other than to fit the experimental data of the
evolution of our universe. Further more, for the case of the phantom fields, it
is in general not even clear how such a field can arise consistently
within quantum field theory.

If, instead of defining the Lagrangian of the field guided  by
the principle  of reproducing the cosmological data, we could obtain
its main properties from a theory intended to describe phenomena
beyond gravity, this would be a step forward to incorporate this
evolution driving mechanism in the scope of a wider theory.

Natural places to look for these fields are supergravity, string
theory, grand  unification theories or other quantum theories of
gravity.

Ideally we would like to find a field which is naturally coupled to
gravity,  or one of its generalizations, ruled by a defined
potential, and see what are its cosmological implications.

Due to the lack of abundance of this kind of scenarios, the choice
for this work will be different, and we will look for this type of
field in the context of the gauge/gravity correspondence
\cite{'t Hooft:1993gx, Susskind:1994vu}, and more in particular, using the
AdS/CFT correspondence \cite{Maldacena:1997re} as it was applied to
a system of D3/D7 branes in \cite{Hovdebo:2005hm}. This analysis provides a
define potential which can be use for the kind of field we are looking for
and an example of a scenario where a phantom field can arise in the context
of a quantum theory.

The AdS/CFT correspondence establishes a duality between string
theory in its supergravity limit and a conformal field theory. The
first is a theory which contains gravity, while the second is a
field theory in a flat background without gravity. This correspondence 
has been extended to more general gauge/gravity situations, and applied 
successfully to obtain physical quantities proper to gauge theories in 
their non-perturbative limit, in the hope of reaching a better 
understanding of non-perturbative QCD \cite{Herzog:2006gh,
Gubser:2006bz,CasalderreySolana:2006rq,Herzog:2006se,Caceres:2006dj,
Caceres:2006as,Chernicoff:2006hi,CaronHuot:2006te,PS,Mateos:2007yp}. See 
\cite{Mateos:2007ay} for a resent review.

By analyzing a particular supergravity configuration in
\cite{Hovdebo:2005hm},  it was possible to determine the spectrum of
mesons, that is, quark-antiquark bound states, in a ${\cal N}$ =2
field theory with fundamental matter. With this result at hand, we compute explicitly the
Lagrangian for the lightest two scalar mesons in this scenario,
which happen to have the same mass. This Lagrangian is valid for
strong gauge coupling in the infrared limit of the field theory, and
we will use this result as a motivation to analyze the consequences
of two mesonic fields, governed by the Lagrangian we
find, if they were present in our universe.

Is necessary for us to mention two reasons why we cannot use this
result as a solid prediction, and then argue why it is still a
strong motivation for the present analysis. On the one hand, the
matter content of the gauge theory, where the mesons studied here live,
is not part of the standard model of particles and has not been observed
to exist in our universe. On the other hand, as mention before, the mesons
we are describing live in a four dimensional Minkowsky space-time,
flat and without gravity, hence not suitable for cosmological
evolution.

Concerning the matter content of the ${\cal N}$=2 theory that we are working with,
we have to say that it is commonly speculated that there could be
in nature a hidden, or dark, sector of particles beyond those included in the
standard model. The particles in this sector could have scape observation for
either of many reasons, like not being in the scope of energies archived
in experiments so far, or simply for being coupled to standard matter
exclusively by gravitational interaction. Assuming the matter
content of the theory at hand to be part of this hidden or dark sector
will permit us, in this work, to perform the analysis we are interested in.

In what respects the lack of gravity in the field theory, we  are
assuming that a similar field to the mesons we find, could be
present in the theory that should describe the processes of
elementary particles in a universe with matter. So we will work
taking a variational approach where the contribution of these mesons to the 
action dictating the cosmological
processes is given by the integral, over the space-time, of the Lagrangian
we find for the fields, multiplied by the determinant of the metric for a
Friedmann-Robertson-Walker universe.

In the next section we will shortly describe the main features of
the supergravity configuration used in \cite{Hovdebo:2005hm}, and
how it is related to the spectrum of the scalar mesons and the
Lagrangian obtained there. In section \ref{Lagr}, we will perform
the explicit calculation of the relevant terms of the Lagrangian in
the infrared limit for the two lightest mesons once they have been
embedded in a FRW metric. Section \ref{potential} will be used to
discus the main features of the Lagrangian found in section
(\ref{Lagr}). The possible cosmological implications will be address
in section \ref{cosmo} and we will finish with some concluding
remarks in section \ref{conclusion}.

\section{Meson spectrum from the supergravity configuration}\label{spectrum}

The original formulation of the AdS/CFT correspondence
\cite{Maldacena:1997re}  establishes a duality between type IIB
string theory in a $AdS_5 \times S^5$ background and a ${\cal N}$ =4
$SU(N_c)$ super Yangs-Mills theory. This background for the string
theory is replacing a stack of $N_c$ D3-branes in the limit where
$N_c >>1$. In this case all the matter fields of the Yang-Mills
theory are in the adjoin representation, since they correspond to
states of open strings starting and ending in a D3-brane, so this
fields come in $N_c$ colors.

In \cite{Karch:2003nh} it was noticed that it is possible to add to
the gauge theory $N_f$ hypermultiplets in the fundamental
representation, which corresponds to have dynamical quarks with
$N_f$ flavors. This is done by starting in Minkowsky space, locating
at the origin a stack of $N_c$ D3-branes extending in the
$x_0,...,x_3$ directions and add a stack of D7-branes extending in
the directions $x_0,...,x_7$. The $N_f$ supermultiplets arise from
the lightest states of open strings going between the D3 and the D7
stacks, and their masses are given by $m_q=L/2\pi\alpha '$, with $L$
the distance between the D3 and D7 branes in the 8-9 plane, and
$\alpha '$ is the square of the string length scale. The insertion
of the D7-branes brakes the supersymmetry from ${\cal N}$=4 to
${\cal N}$=2. In the limit where $g_sN_c>>1$, with $g_s$ the string
coupling, it is possible to replace the D3-branes for an $AdS_5
\times S^5$ background, and further more, as noticed in
\cite{Skenderis:2002vf}, in the regime where $N_c>>N_f$, the
D7-branes can be considered probes over this space.

The gauge theory dual to the supergravity configuration previously
describe has a rich spectrum of mesons, that is, quark-antiquark
bounded states. The precise masses for this spectrum can be computed
\cite{Hovdebo:2005hm} by varying the Born-Infeld action for the
fields over the D7-branes.

The mesons of interest for this work will be part of the open string
excitations of the D7-branes, corresponding to scalar and gauge
fields, which dynamic is described by the action
\cite{Polchinski:1998rq}
\be
S_{D7}=-\mu_7 \int
\sqrt{-det(P[G_{ab}]+2\pi \alpha ' F_{ab})}+ \frac{(2\pi\alpha
')^2}{2} \mu_7 \int P[C^{(4)}]\wedge F \wedge F, \label{SgenD7}
\ee
where the integral has to be performed over the eight dimensional
volume of the D7-brane and $P[X]$ stands for the pullback of $X$ to
this volume. The brane has a tension given by
$\mu_7=[(2\pi)^7g_s\alpha'^4]^{-1}$. The metric $G_{ab}$ is that of
the space $AdS_5 \times S^5$, where the D7-branes are embedded, and
it is given by
 \be
 ds^2=\frac{r^2}{R^2}\eta_{\mu\nu}dx^\mu dx^\nu +
\frac{R^2}{r^2}(dx^i\cdot dx^i), \label{dsads5}
\ee
with
$r^2=x^i\cdot x^i$ and $R=\sqrt{4\pi g_s N_c}\alpha'$. The indexes
$\mu$ and $\nu$ run from 0 to 3 and are contracted with the
Minkowsky metric $\eta_{\mu\nu}$, while the $i$'s run from 4 to 9.
The four-form in the Wess-Zumino term is
\be
C^{(4)}=\frac{r^4}{R^4}dx^0\wedge dx^1\wedge dx^2\wedge dx^3.
\label{C4}
\ee
The $F$ entering action (\ref{SgenD7}) is the field
strength of the gauge field $A$ living over the D7-branes.

If we locate the D7-branes at a distance $L$ from the D3-branes in
the 8-9 plane, which without lost of generality can be done by
fixing its coordinates $X^8=0$ and $X^9=L$, the induced metric can
be written as
\be
ds^2=\frac{\rho^2+L^2}{R^2}\eta_{\mu\nu}dx^\mu
dx^\nu + \frac{R^2}{\rho^2+L^2}(d\rho^2+\rho^2d\Omega_3^2),
\label{dsD7}
\ee
where $\rho$ and $\Omega_3$ are spherical
coordinates in the 4567 space, with $\rho$ realted to $r$ by
$\rho^2=r^2-L^2$.

In the case $L=0$ this metric reduces to that of $AdS_5\times S^3$,
so the $AdS$ factor suggests the dual theory to still be conformally
invariant, and it is indeed the case. This conformal invariance is
also clear in the gravity side, since the metric (\ref{dsD7}) with
$L=0$ is invariant under the simultaneous rescaling of
$x^\mu\rightarrow\Lambda x^\mu$, and $\rho\rightarrow\rho/\Lambda$.
The coordinates $x^\mu$ are the ones associated to the space of the
gauge theory, so from this last remark we see that the physics of
small distances, or high energies, in the dual theory,
corresponds to large values of $\rho$, and conversely, the long
distance or low energy physics, corresponds to small values of
$\rho$. When $L\neq 0$ the gauge theory is not conformally
invariant, since the value of $L$ introduces a scale which, as we
saw, is proportional to the mass of the fundamental matter. It can
be proved that the gauge theory recovers approximately the conformal
invariance if $\rho >> L$, that is, in the ultraviolet limit, where
the typical energy of the processes is much bigger than $m_q$.
Consistently the metric (\ref{dsD7}) with $L\neq 0$ is not invariant
under the rescaling mentioned above, but it recovers this as an
approximated symmetry if $\rho >> L$.

We will be interested in the mesons coming from the scalar fields,
which can be represented by the fluctuations $\phi$ and $\varphi$
around the fiducial embedding just presented,
 \be
 X^8=0+2\pi \alpha
' \phi \,\,\, , \,\,\, X^9=L+2\pi \alpha ' \varphi.
\label{campos}
\ee
To study these fields we can consistently set $A=0$, in which
case the Lagrangian density to third order in the fields is
\cite{Hovdebo:2005hm} \be {\cal L}\simeq -\mu_7
\sqrt{-det(g_{ab})}(1+2(R\pi\alpha
')^2\frac{g_{cd}}{r^2}(\partial_c\phi\partial_d\phi+\partial_c\varphi\partial_d\varphi)),\label{Saprox}
\ee where $g_{ab}$ is the metric given by (\ref{dsD7}) and
$g_{cd}/r^2$ is to be expanded only to first order in the fields.
The determinant of the induce metric entering this Lagrangian
factorizes in such a way that it is independent of the fields (\ref{campos}), so
(\ref{Saprox}) is the correct Lagrangian to third order
\cite{Hovdebo:2005hm}. In the next section it will be clear why
taking the Lagrangian only to third order is a good approximation in
the infrared limit.

Writing explicitly $g_{bc}/r^2$ to linear order in the fields we get
\bea
{\cal L}\!\simeq\!\!\!\!\! &-&\!\!\!\!\!\mu_7(2\pi\alpha')^2\rho^3
\sqrt{\tilde{g}}\left(\frac{R^4}{2(\rho^2+L^2)^2}\eta^{\mu\nu}
(\partial_\mu\phi\partial_\nu\phi+\partial_\mu\varphi\partial_\nu\varphi)+
\frac{1}{2}(\partial_\rho\phi)^2+\frac{1}{2}(\partial_\rho\varphi)^2 \right.
\nonumber \\
&+&\left.\frac{1}{2\rho^2}\tilde{g}^{ij}(\partial_i\phi\partial_j\phi+
\partial_i\varphi\partial_j\varphi)-\frac{4R^4L\pi\alpha'}{(\rho^2+L^2)^3}
\eta^{\mu\nu}\varphi(\partial_\mu\phi\partial_\nu\phi+\partial_\mu\varphi
\partial_\nu\varphi)\right), \label{Laprox}
\eea
where $\tilde{g}^{ij}$ is the metric over the three sphere.

To take a look at the spectrum of the mesons described by $\phi$ and
$\varphi$,  let's work for a moment to second order in
(\ref{Saprox}), which is equivalent to ignoring the dependence of
$g_{cd}/r^2$ on the fields. We see that the equation of motion for
either of the fields is
 \be
\frac{R^4}{(\rho^2+L^2)^2}\eta^{\mu\nu}\partial_\mu\partial_\nu \Psi
+ \frac{1}{\rho^3}\partial_\rho (\rho^3\partial_\rho \Psi)
+\frac{1}{\rho^2}\tilde{g}^{ij}\nabla_i\nabla_j \Psi=0, \label{em2}
\ee where $\nabla_i$ is the covariant derivative in the sphere and
$\Psi$ stands for any of the two fields.

To solve this equation is possible \cite{Hovdebo:2005hm} to use separation of variables and write
\be
\Psi=\psi(\rho)e^{ik\cdot x}{\mathcal{Y}}^{\ell}(S^3), \label{separacion}
\ee
with ${\mathcal{Y}}^{\ell}(S^3)$ the spherical harmonics in $S^3$, satisfying
\be
\tilde{g}^{ij}\nabla_i\nabla_j {\cal{Y}}^{\ell}=-{\ell}({\ell}+2){\cal{Y}}^{\ell}. \label{SHar}
\ee

Given (\ref{em2}) and (\ref{separacion}), the solution for
$\psi(\rho)$ is \be
\psi(\rho)=\rho^{\ell}(\rho^2+L^2)^{-\alpha}F(-\alpha,-\alpha+{\ell}+1;{\ell}+2;-\rho^2/L^2),
\label{psi} \ee where $F$ is the standard hypergeometric function
and \be
\alpha=\frac{1}{2}\left(-1+\sqrt{1-\frac{k^2R^4}{L^2}}\right).
\label{alpha} \ee The mesons described by the gauge/gravity
correspondence are associated  to the normalizable modes of the
fields over the D7-brane. In this case, for $\Psi$ to be
normalizable, it is necessary that $\psi(\rho)$ remains finite as
$\rho\rightarrow\infty$, condition that will be satisfied only if
\be -\alpha+{\ell}+1=-n, \,\,\,\,\,\,\,\,\,\,\,\,\,\, {\mathrm{for}}
\,\,\,\,\,\,\,\,\,\,\,\,\,\,\, n=0,1,2,3... \label{quantC} \ee so
given (\ref{alpha}) and that the mass of the fields from the four
dimensional perspective is given in terms of the momentum $k$ in
(\ref{separacion}) by $M^2=-\eta^{\mu\nu}k_\mu k_\nu$, the spectrum
of the mesons is found to be \be
M_{n,{\ell}}=\frac{2L}{R^2}\sqrt{(n+{\ell}+1)(n+{\ell}+2)}.
\label{espectro} \ee

To consider now the Lagrangian to third order on the fields
(\ref{Laprox}), we can decompose each field in its four dimensional
spectrum with the same dependence on $\rho$ and the angular part as
for the quadratic case, so we write
\be
 \phi=\sum_a
\psi_{n_a{\ell}_a}(\rho){\cal{Y}}^{{\ell}_a}(S^3)\tilde{\phi}(x^\mu)
\,\,\,\,\, {\mathrm{and}}\,\,\,\,\, \varphi=\sum_b
\psi_{n_b{\ell}_b}(\rho){\cal{Y}}^{{\ell}_b}(S^3)\tilde{\varphi}(x^\mu),
\label{expcamp} \ee where the $a$'s and $b$'s run over all the
values of $n, {\ell}$, and also include the other quantum numbers
proper to the spherical harmonics.

Now, what is left for us to do is to find the four dimensional
Lagrangian  for the lightest mesons.

\section{The four dimensional Lagrangian in the infrared limit}\label{Lagr}

In the infrared limit of the theory, we would like to consider the
lightest  scalar mesons of the spectrum (\ref{espectro}), which are
those with quantum numbers $n=\ell=0$, corresponding to the
solutions \be
\phi_0=\frac{1}{\sqrt{2}\pi(\rho^2+L^2)}\tilde{\phi}(x)
\,\,\,\,\,\,\,\,\,\,\,\,\,\,\,\,\,\,\,\,\,\,\,\, {\mathrm{and}}
\,\,\,\,\,\,\,\,\,\,\,\,\,\,\,\,\,\,\,\,\,\,\,\,
\varphi_0=\frac{1}{\sqrt{2}\pi(\rho^2+L^2)}\tilde{\varphi}(x),
\label{campos4} \ee with mass $M_0=2^{3/2}L/R^2$.

To obtain the four dimensional Lagrangian, we have to substitute
(\ref{campos4}) in to (\ref{Laprox}) and integrate over the $\rho$
direction and the three sphere. The result of doing so is \bea
{\cal L}_4\simeq &-&\mu_7(2\pi \alpha')^22\pi^2 \left[ \frac{R^4}
{24L^4}(\partial_\mu\tilde{\phi}\partial^\mu\tilde{\phi}+
\partial_\mu\tilde{\varphi}\partial^\mu\tilde{\varphi}) \right. \nonumber \\
&+& \left. \frac{4}{12L^2}(\tilde{\phi}^2+\tilde{\varphi}^2)-\frac{R^4\alpha'}
{ 10\sqrt{2}L^7}\tilde{\varphi}(\partial_\mu\tilde{\phi}\partial^\mu\tilde{\phi}
+\partial_\mu\tilde{\varphi}\partial^\mu\tilde{\varphi})\right],
\eea
or cleaning up a little bit
\be
{\cal L}_4\simeq A\left[ \frac{1}{2}(1-B\tilde{\varphi})(\partial_\mu\tilde{\phi}
\partial^\mu\tilde{\phi}+\partial_\mu\tilde{\varphi}\partial^\mu\tilde{\varphi})+
\frac{1}{2}{M_0}^2(\tilde{\phi}^2+\tilde{\varphi}^2)\right] ,
\label{Laprox4d} \ee with \be
A=-\mu_7(2\pi\alpha')^22\pi^2\frac{R^4}{12L^4}
\,\,\,\,\,\,\,\,\,\,\,\,\,\,\,\,\,\,\,\,\,\,\,\,\, {\mathrm{and}}
\,\,\,\,\,\,\,\,\,\,\,\,\,\,\,\,\,\,\,\,\,\,\,\,\,\,\,\,\,\,\,
B=\frac{6\sqrt{2}\alpha'}{ 5L^3} . \label{Langcons} \ee To get a
further simpler expression we can make a field redefinition and
take $\hat{\phi}=B \tilde{\phi}$ and $\hat{\varphi}=B
\tilde{\varphi}$ to finally get \be {\cal L}_4\simeq {\cal
N}_4\left[
\frac{1}{2}(1-\hat{\varphi})(\partial_\mu\hat{\phi}\partial^\mu\hat{\phi}+
\partial_\mu\hat{\varphi}\partial^\mu\hat{\varphi})+\frac{1}{2}{M_0}^2(\hat{\phi}^2+\hat{\varphi}^2)\right]
, \label{Laprox4dsim} \ee where \be {\cal
N}_4=\frac{A}{B^2}=\frac{-25}{108}\mu_7\pi^4L^2R^4. \label{norm4}
\ee Before continuing with the calculation, now we are in position
to notice why the Lagrangian to third order in the fields is a
good approximation in the infrared limit. As mention in the previous
section, the infrared limit of  the gauge theory is given by $\rho
<< L$, and so, according to (\ref{campos4}) we can approximate \be
\phi_0\simeq\tilde{\phi}(\sqrt{2}\pi L^2)^{-1}\,\,\,\,\,\,\,\,\,
{\mathrm{and}}\,\,\,\,\,\,\,\,\,
\varphi_0\simeq\tilde{\varphi}(\sqrt{2}\pi L^2)^{-1}.
\label{fieldsIRlim} \ee In the gravity side $2\pi\alpha'\phi_0$ and
$2\pi\alpha'\varphi_0$ are fluctuations of the position of the
D7-brane around the fiducial embedding, so they should be small
compared to $L$, and as a consequence of (\ref{fieldsIRlim}) the
inequalities $\sqrt{2}\alpha'\tilde{\phi}<< L^3$ and
$\sqrt{2}\alpha'\tilde{\varphi}<< L^3$ should hold. From the
definitions of $\hat{\phi}$ and $\hat{\varphi}$ we get to the
conclusion that these last fields have to satisfy $\hat{\phi}<<6/5$
and $\hat{\varphi}<<6/5$. Given these numbers, we see that the fields
$\hat{\phi}$ and $\hat{\varphi}$ have to be small, but not
neglectable, compare to the unit when working in the normalization
appropriated for (\ref{Laprox4dsim}). Terms of higher order in the
fields would then be suppressed in the infrared limit, but it is
still of consequence to analyze the Lagrangian to third order.
Notice that this is not necessarily the case in the ultraviolet
regime $\rho\rightarrow\infty$, since the values of $\tilde{\phi}$
and $\tilde{\varphi}$ are not bounded in (\ref{campos4}) by the
inequalities $2\pi\alpha'\phi_0<<L$ and $2\pi\alpha'\varphi_0<<L$
given that the factor $(\rho^2+L^2)^{-1}$ can be arbitrarily small.
We also know that in this regime modes of higher mass would be
excited, and it can be seen that also for these modes the behavior of $\Psi(\rho)$ when
$\rho\rightarrow\infty$, does not permit to place a bound on the
value of the four dimensional fields, so we are confronted with the
fact that to study the ultraviolet limit of the theory, it is
necessary to include all the tower of massive states and to take the
Lagrangian to higher orders in the fields.

Let's continue the calculation relevant for our current analysis.
Now  we have seen that expression (\ref{Laprox4dsim}) is the
appropriated Lagrangian for the scalar mesons, in four dimensional
Minkowsky space, in the infrared regime of the gauge theory dual to
the supergravity configuration that we have been studding so far.
This is what will suggest our starting point for the rest of the
analysis, which will consist on considering two fields described by
the Lagrangian (\ref{Laprox4dsim}), but instead of living in a flat
space, we will think of them living in a FRW universe, and explore
the behavior of such a system.

The action of relevance is
\be
S=\int {\cal N}_4 a^3(t)\left[
\frac{1}{2}(1-\hat{\varphi})
(\partial_\mu\hat{\phi}\partial^\mu\hat{\phi}+\partial_\mu
\hat{\varphi}\partial^\mu\hat{\varphi})+\frac{1}{2}{M_0}^2(\hat{\phi}^2+\hat{\varphi}^2)\right]
. \label{SFRW}
\ee
Something important to notice here is that the sign of the kinetic energy in eqs.(\ref{SFRW}) depends on the
value of $\hat\vp$. If $\hat\vp$ is smaller than one then the kinetic
terms are positive but for $\hat\vp>1$ the kinetic terms become negative
and the fields could be phantom fields. We argued above that in the dual theory strictly derived from the supergravity configuration, $\hat\vp$ cannot be bigger than the unit. Nonetheless, we are now analyzing the Lagrangian, originally  obtained from the gauge/gravity correspondence, in the context of cosmic evolution. Given the relevance of phantom fields in this scenario, and how unclear it is the way this type of field can arise from a quantum theory, we will permit our selfs to explore also this range of values for $\hat\vp$ in a speculative fashion. With a naive inspection of
eq.(\ref{SFRW}) we could think that
the regions with $\hat\vp>1$ and $\hat\vp<1$ could  be dynamically
connected, however, as we will see below, this is not the case in this model and therefore, it does not allow a cross over the phantom line.

Working in the approximation of an homogeneous field we have
$\partial_i\Psi \ll  \dot\Psi$, where the variation of the fields in the spacial
directions  is very small in comparison with the variation
in the time direction and the dot stands for the derivative with respect of time,
the action (\ref{SFRW}) leads to the
following equations of motion for the fields,
\bea
\ddot{\hat{\phi}}(1-\hat{\varphi})+3H \dot{\hat{\phi}}(1-\hat{\varphi})
-\dot{\hat{\varphi}} \dot{\hat{\phi}}+ {M_0}^2\hat{\phi} &=&0, \label{emphi} \\
\ddot{\hat{\varphi}}(1-\hat{\varphi})+3H \dot{\hat{\varphi}}(1-\hat{\varphi})+
 \frac{1}{2}(\dot{\hat{\phi}}^2-  \dot{\hat{\varphi}}^2)+{M_0}^2\hat{\varphi}&=&0,
\label{emvarphi}
\eea
where $H =\dot{a}/a$ is Hubble's constant.

\section{Characteristics of the equations of motion}\label{potential}

To extract the physics described by eq.(\ref{SFRW}) we would like to
first find a subfamily of the solutions to this equations which
permits a simple analysis of the general behavior. With this in
mind, we could   set $\hat{\varphi} $ or $\hat\phi$ constant. Let us
first set $\hat\vp$ constant. In this case eq.(\ref{emphi}) becomes
that of a standard scalar field with a constant mass $m^2\equiv
M_0^2/(1-\hat\vp)$,
\be
\ddot{{\hat\phi}}+3H \dot{{\hat\phi}}=- m^2\hat\phi \equiv - \fr{dB}{d\hat\phi},
\ee
and a potential $B(\phi)=m^2\hat\phi^2/2$. If   $m^2$ is
positive, i.e. $\hat\vp<1$, then the field $\hat\phi$ oscillates
around the origin with decreasing amplitude and its  energy density
$\rp$ redshifts as matter with $w_{\hat{\phi}}=0 $ and $\rho_{\hat\phi}\propto
a^{-3}$. On the other hand if $m^2$ is negative then the $\hat\phi$
is unstable and grows  to infinity. In this case we see from
eq.(\ref{SFRW}) that
 the   kinetic energy is negative  and $\hat\phi$  behaves  as a phantom field.

Now, considering  the field $\hat\vp$ it is useful to redefine it in
order to obtain a standard canonical term $L_k=\pu \vp \pU\vp/2$,
via\footnote{This new redefinition of $\vp$ should not be confused
with that in eq.(\ref{campos}).}
\be\la{dvp} \pu \vp=
\sqrt{\N(1-\hat\vp)}\;\pu\hat\vp.
\ee
Integrating eq.(\ref{dvp}) we
get a canonical scalar field \be\la{2vp} \vp= - q
(1-\hat\vp)^{3/2},\hspace{1cm}\hat\vp=1- \le(\fr{\vp}{q}\ri)^{2/3}
\ee with $q\equiv 2\sqrt{\N}/3$.   For   $1-\hat\vp>0$  the field
$\vp$ is real while for $1-\hat\vp<0$ it becomes imaginary. We can
write both regions in a single notation in terms of  a real scalar
field $\vp$ by introducing a parameter $\xi$ which takes the value
$\xi=1$ for $1-\hat\vp>0$ and $\xi=-1$ for $1-\hat\vp<0$.
Furthermore since the Lagrangian is symmetric between $\vp
\leftrightarrow -\vp$ we will take the positive sign, without loss
of generality, and define the canonical scalar field as
\be\la{3vp}
 \vp= q\,
\,[\xi(1-\hat\vp)]^{3/2},\hspace{1cm}\hat\vp=1-\xi\le(\fr{\vp}{q}\ri)^{2/3}
\ee and the kinetic energy  becomes \be
L_k=\fr{\N}{2}(1-\hat\vp)\partial_\mu\hat\vp\pU\hat\vp =\frac{\xi
}{2}\partial_\mu\vp\pU\vp,
\ee
showing that the sign of the kinetic term
can be positive or negative depending on the value of $\xi$, and the potential is \be\la{V} V(\vp)=
\fr{M_0^2}{2}\le(1- \xi\le(\fr{\vp}{q}\ri)^{2/3}\ri)^2. \ee Using
eq.(\ref{3vp}) and defining\footnote{This new redefintion of $\phi$
should not be confused with that in eq.(\ref{campos}).}
$\phi=\sqrt{\N} \hat\phi$ the Lagrangian in eq.(\ref{SFRW}) becomes
\be\la{s2}
 S=\int  a^3(t)\left[\frac{\xi}{2}\partial_\mu\vp\pU\vp +
\frac{\xi}{2}\partial_\mu\phi \partial^\mu{\phi}
\le(\fr{\vp}{q}\ri)^{2/3} + V(\vp) + \frac{1}{2}{M_0}^2
\phi^2\right]
\ee
with   the potential $V(\vp)$ defined in
eq.(\ref{V})  and $\xi=\pm 1$ depending on the initial sign of
$1-\hat\vp$. The equations of motion  from eq.(\ref{s2}) are
\bea\la{eqM}
\xi \ddot{{\phi}}+3H
\xi\dot{{\phi}}+\fr{2\xi}{3q^{2/3}}
\fr{\dot\phi\dot\vp}{\vp^{1/3}}\le(\fr{\vp}{q} \ri)^{-2/3}
+  M_0^2\phi \le(\fr{\vp}{q} \ri)^{-2/3} &=&0 \non \\
\xi\ddot{{\vp}}+3H\xi\dot\vp-\fr{\xi \dot\phi^2 }{3q^{2/3}\vp^{1/3}} +  \fr{dV}{d\vp}  & = & 0
\eea
with $dV/d\vp= 2\xi M_0^2(1- \xi(\vp/q)^{2/3})/3q^{2/3}\vp^{1/3}$.

The  potentials $V(\vp)$ and $B(\phi)\equiv M_0^2\phi^2/2$  are plotted in fig.1.
For $\xi=1$ the potential $V$  has a minimum at
$\vp_o=q=\pm 2\sqrt{\N}/3$ (i.e. for $\hat\vp=0$)
and $B$ at $\phi_o=0$. The field
derivative of $V$  at $\vp=0$ (i.e. for $\hat\vp=1$) diverges. The steep potential
at $\vp=0$ signals that we cannot cross over the
value of $\vp=0 $ (i.e. from regions with $\hat\vp>1$ to $\hat\vp<1$).
The dynamical evolution of $\vp$ is to roll down the potential $V$ to its
minimum. Expanding $V$ around the minimum and keeping the leading term only we have
$V=2 M_0^2(\vp-q)^2/9 q^2 $. The fact that $V$ is quadratic around the minimum
implies that the energy density of $\vp$ redshifts as matter, i.e.
$\rho(\vp)\propto a^{-3}$, and has a mass $m^2=d^2V/d\vp^2=4 M_0^2/9q^2=M_0^2/3\N$.
At the minimum of the potential $V(\vp)$ one has $\vp=\vp_o=q=\pm 2\sqrt{\N}/3$
and the kinetic term for the field $\phi$ in eq.(\ref{eqM}) becomes canonically normalized,
i.e. $ L_k = \partial_\mu\phi \partial^\mu{\phi}/2$.
We find a similar behavior  for  $\phi$  where the field oscillates around $ \phi_o=0$,
which is the minimum of the potential $B=M_0^2\phi^2/2$, and has a mass $M_0$.
Both fields $\vp$ and $\phi$ oscillate around the minimum of their potential, however,
the mass of these fields  differs. The mass of $\phi$ is $M_0$ while
that of $\vp$ is $m=M_0/\sqrt{3\N}$ and $M_0^2/m^2=3\N$. Therefore
for a small $\N$, i.e. $\N\ll 1$,  one has $M_0\ll m$.

Now, if $\xi=-1$ then the potential $V$ becomes
$V= M_0^2(1+(\vp/q)^{2/3})^2/2$ and it has a minimum
value at $\vp=0$. However, since in this case the kinetic
term is negative the evolution of $\vp$ is to grow
to $\vp \rightarrow \infty$.
The field $\phi$ has  also a runaway behavior $\phi \rightarrow \pm \infty$ due to the negative
kinetic energy. This behavior is equivalent to
having a scalar field with positive kinetic term and
with a negative unbounded potential $\hat{V}=-V=-M_0^2(1+(\vp/q)^{2/3})^2/2$
and $\hat{B}=-B=-M_0^2\phi^2/2$.

As we have seen in both cases $\xi=1$ and $\xi=-1$ the point
$\vp=0$, i.e. $\hat\vp=1$, is a repulsive point and $\vp$
runs away from it. For positive kinetic term the fields
reach a minimum value while for negative kinetic terms they
grow to infinity and we have no dynamical  cross over the point $\vp=0$.

\begin{figure}[htp!]
\begin{center}
\includegraphics[width=6.5cm]{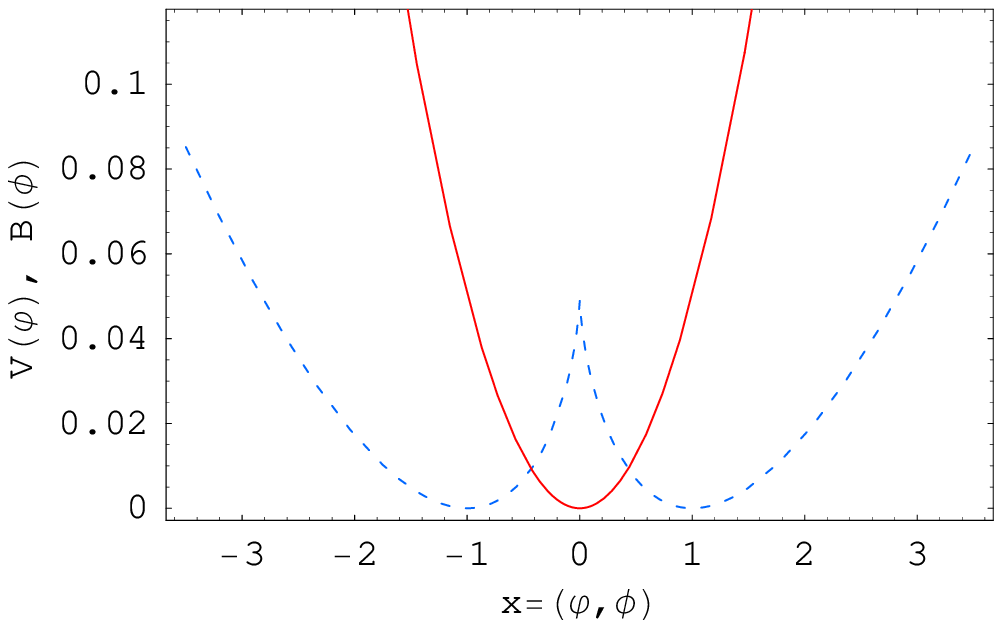}
\includegraphics[width=6.5cm]{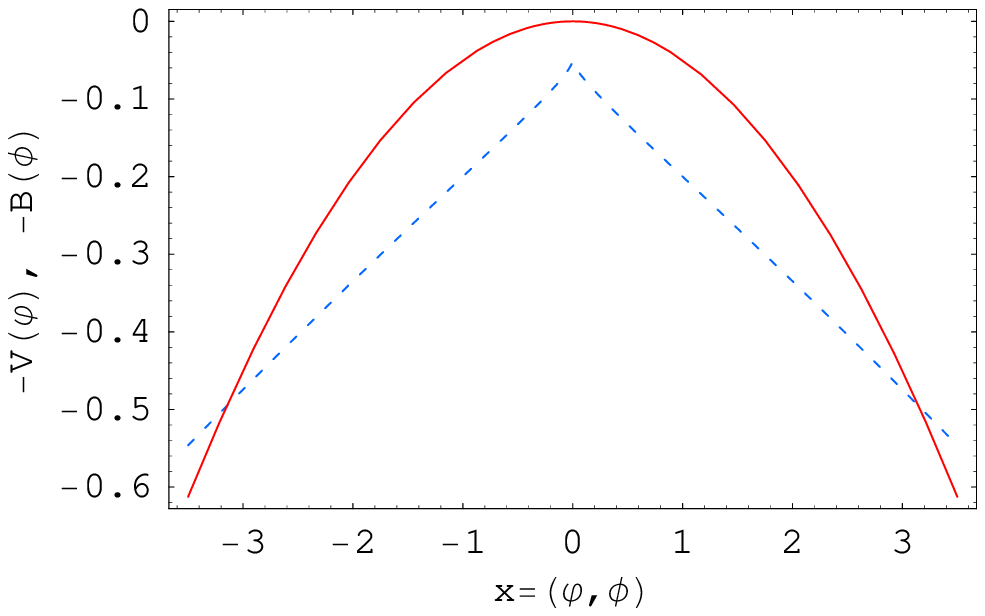}
\end{center}
\caption{\small{We show the potential $V(\vp)$ (c.f. eq.(\ref{V}))
and $B=M_0^2\phi^2/2$ (blue (dotted), red (solid) respectively) for
$M_0^2= 0.1, q=1$ and $\xi=1$ (left figure). The potentials are
symmetric around $\vp=0, \phi=0$ and it is clear that the derivative
of $V$ is discontinuous and diverges at $\vp=0$. Notice that the
potentials have different curvature  at their minimum i.e. different
mass. We show the potential $V(\vp)$ and $B=M^2\phi^2/2$ (blue
(dotted), red (solid) respectively) for $M^2=0.1, q=1$ but for
$\xi=-1$ (right figure). We have plotted the negative potentials
since the dynamics of the fields is to roll down  $-V$ and $-B$, due to the
negative kinetic energy. }} \la{fig1}
\end{figure}

\begin{figure}[htp!]
\begin{center}
\includegraphics[width=6.5cm]{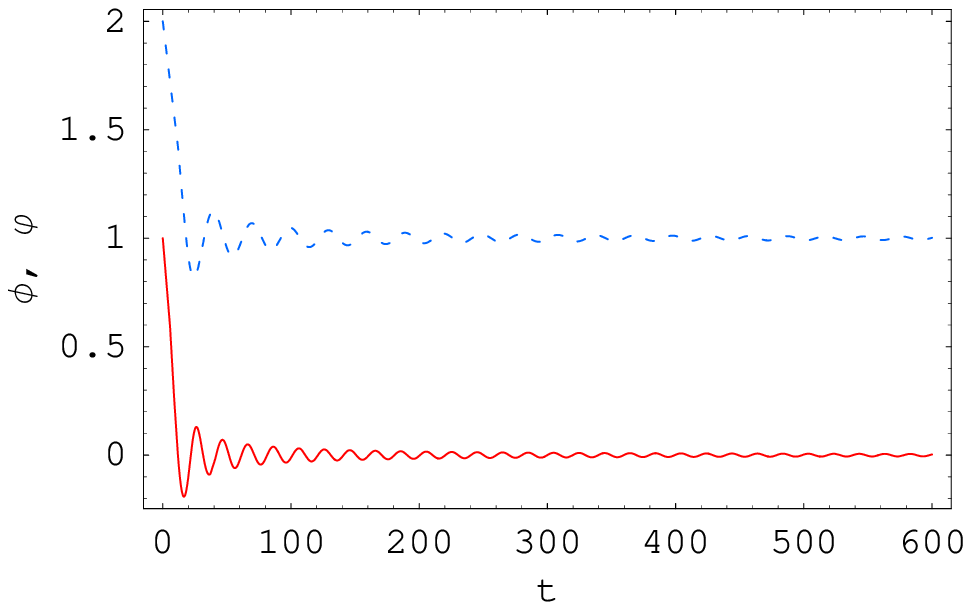}
\includegraphics[width=6.5cm]{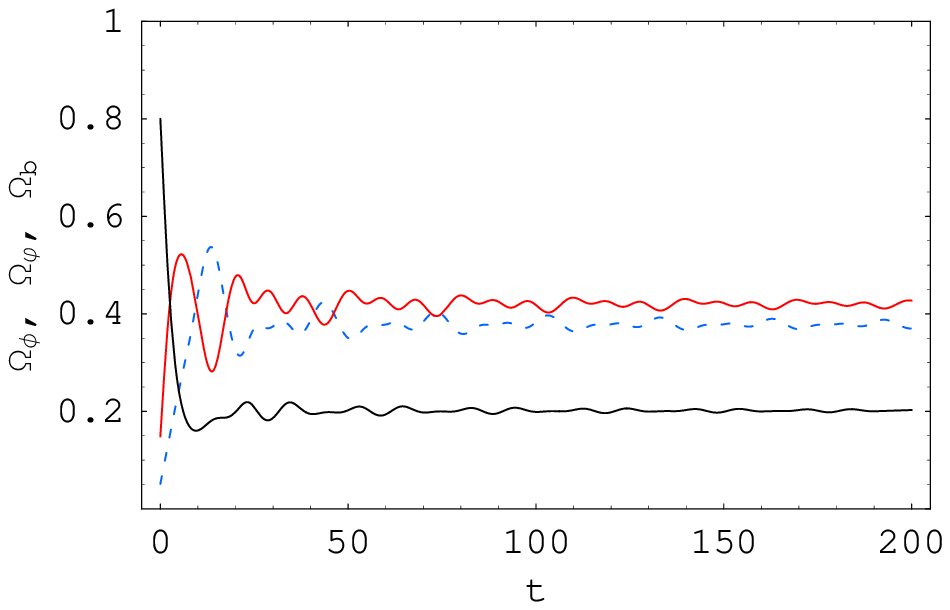}
\includegraphics[width=6.5cm]{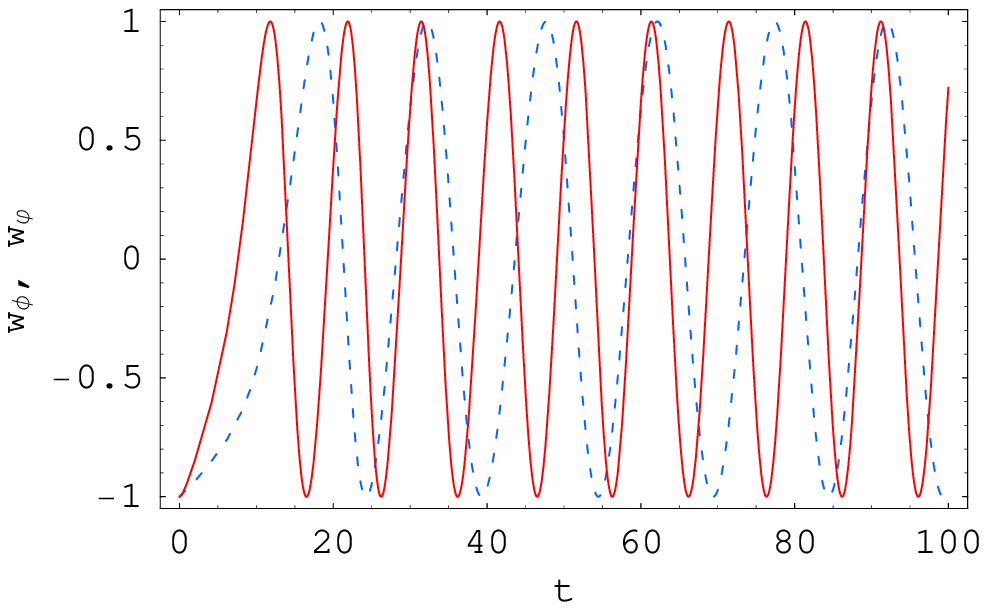}
\includegraphics[width=6.5cm]{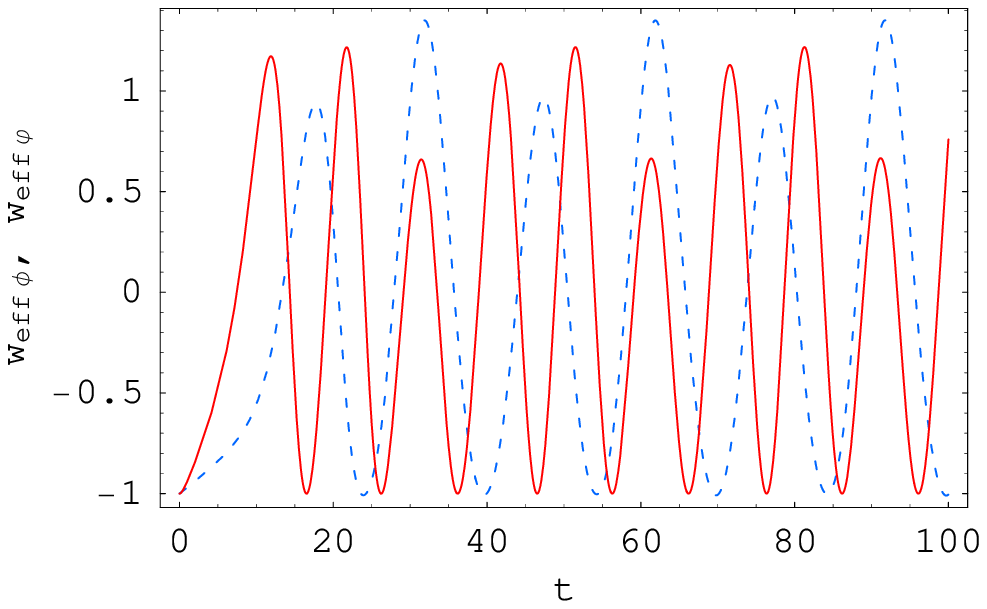}
\end{center}
\caption{\small{ We show for $M_0^2= 0.1, q=1$ and $\xi=1$ the
evolution of $\vp$ and $\phi$ (blue (dotted), red (solid)
respectively), $\Omvp,\Omp,\Om$ (blue (solid), red (dotted) and
black respectively)
 and the equation of state $\wvp, w_\phi$ (blue (dotted), red (solid) ) and
 $\wvpe, \wpe$ (blue (dotted), red (solid)). The effect of the interaction term $\Gamma$ is small.
 The redshift of $\rp,\rvp$ is that of matter $w=0$, i.e. $\rp,\rvp\propto 1/a^{3}$.
 The initial value for matter is $\Om=0.8$. }}
\la{fig2}
\end{figure}

\begin{figure}[htp!]
\begin{center}
\includegraphics[width=6.5cm]{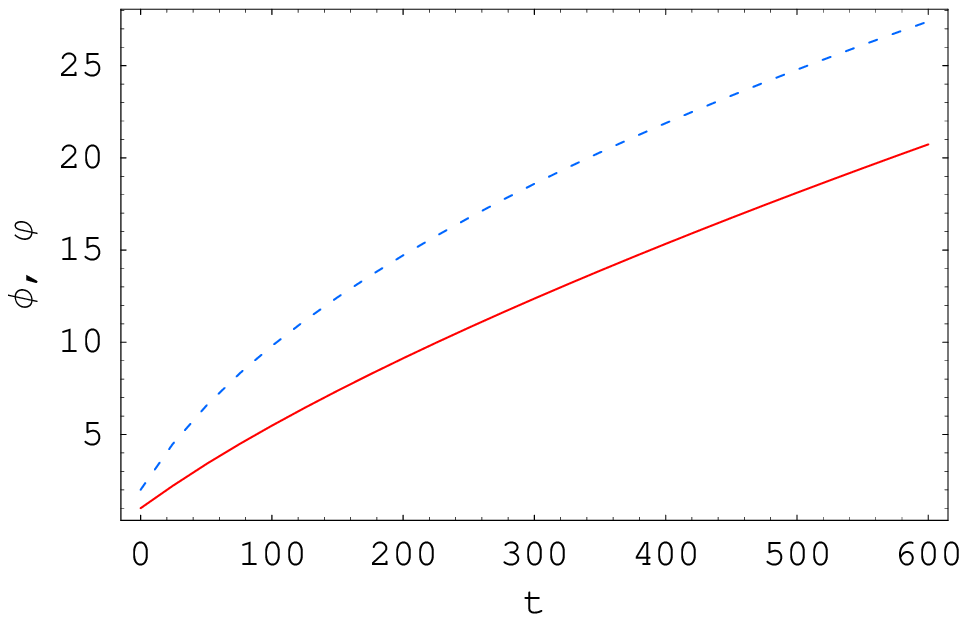}
\includegraphics[width=6.5cm]{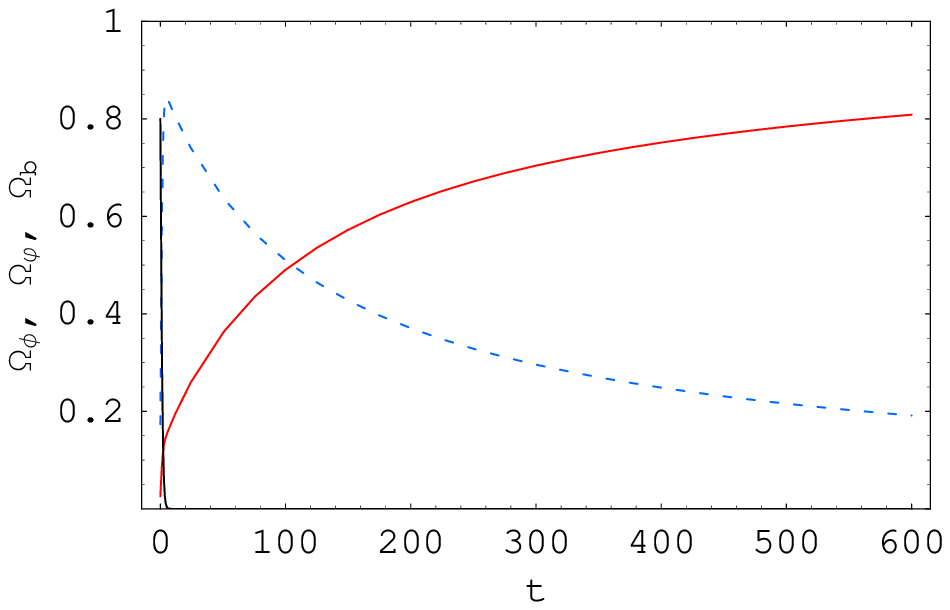}
\includegraphics[width=6.5cm]{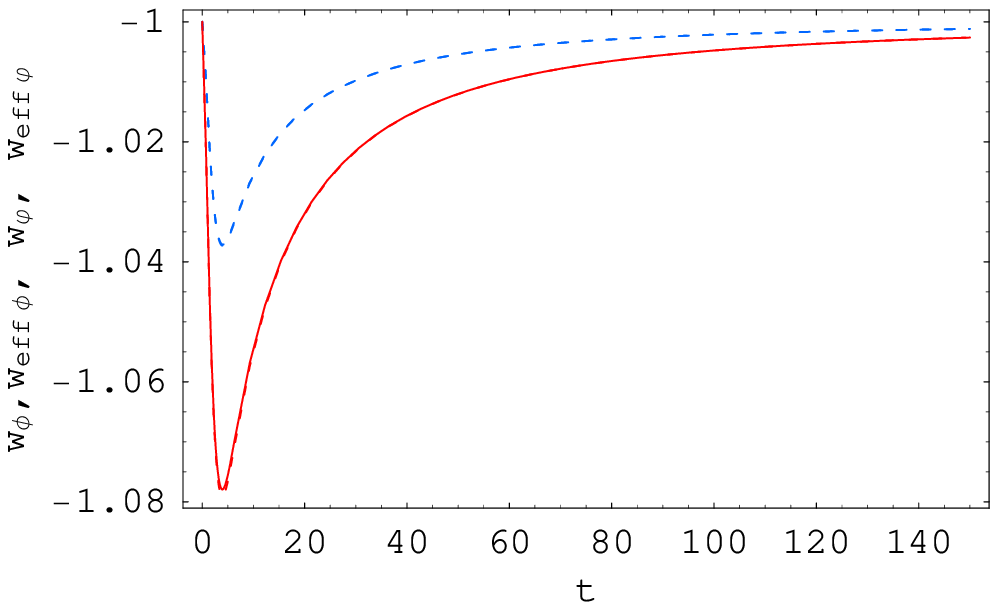}
\end{center}
\caption{\small{ We show for $M_0^2= 0.1, q=1$ and $\xi=-1$ the
evolution of $\vp$ and $\phi$ (blue (dotted), red (solid)
respectively), $\Omvp,\Omp,\Om$ (blue (dotted), red (solid)  and
black respectively)
 and the equation of state $\wvpe, \wpe$ (blue (dotted), red (solid) ).
  The effect of the interaction term  $\Gamma$ is again  very small
and  it cannot be distinguish in the graphs for the $w_\vp$ and
$\wvpe$ or $w_\phi$ from $\wpe$. For both fields the equation of state is smaller than -1 at
all times and they dominate the universe. The initial value for
matter is $\Om=0.8$ and dilutes so fast that it can hardly  be seen
in the graph.    }} \la{fig3}
\end{figure}

\section{Cosmological implications}\label{cosmo}

If we want to study the cosmological evolution of the fields $\vp, \phi$
it is useful to define the energy density and pressure.
The equation of motions in (\ref{eqM}) can be written in
terms of the total energy density and pressure defined  by
\bea\la{rp}
\rho&=&\fr{\xi}{2} \dot\vp^2+\fr{\xi}{2} \le(\fr{\vp}{q}\ri)^{2/3}  \dot\phi^2 +V(\vp)+ \fr{1}{2}M_0^2\phi^2 \\
p&=&\fr{\xi}{2} \dot\vp^2+\fr{\xi}{2} \le(\fr{\vp}{q}\ri)^{2/3}  \dot\phi^2 -V(\vp)- \fr{1}{2}M_0^2\phi^2
\eea
and the energy continuity equation $T^\mu\;_0;\mu=0$ gives the
usual cosmological evolution \be\la{dr} \dot\rho= -3H(\rho+p) \ee
where $H$ is the Hubble constant given by $H^2=8\pi \rho/3$
(here we are taking $ G\equiv 1$ and we assume a flat space). In order to analyze the energy density contribution
from each field we define
\be\la{rvp}
\rvp\equiv \fr{\xi}{2} \dot\vp^2 +V
 ,\hspace{1cm} p_\vp\equiv \fr{\xi}{2} \dot\vp^2 - V
 \ee
and
\be\la{rp2}
\rp\equiv \fr{\xi}{2} \le(\fr{\vp}{q}\ri)^{2/3}  \dot\phi^2 +  \fr{1}{2}M_0^2\phi^2,\hspace{1cm}
p_\phi\equiv \fr{\xi}{2} \le(\fr{\vp}{q}\ri)^{2/3}  \dot\phi^2  - \fr{1}{2}M_0^2\phi^2
 \ee
 with $\rho=\rvp+\rp$ and $p=p_\vp+p_\phi$. With these variables the equation of
motions are
 \bea\la{drp}
\dot\rvp + 3H(\rvp+p_\vp)&=& \Gamma\\
\dot\rp + 3H(\rp+p_\phi) &=& -\Gamma \la{drvp}
\eea
with
\be \Gamma= \fr{\xi\dot\phi^2\dot\vp}{3q^{3/2}\vp^{1/3}}
 \ee
 Clearly if we sum eq.(\ref{drp})
and (\ref{drvp}) we recover eq.(\ref{dr}) and also the equation of motions (\ref{eqM}). The interaction between
$\phi$ and $\vp$ given by the term in the Lagrangian $\vp^{2/3}
\dot\phi^2$ is reflected via the $\Gamma$ term in eqs.(\ref{drp})
and (\ref{drvp}). The sign of $\Gamma$ depends only on $\xi\dot\vp$.

The usual equation of state parameters are defined as
\be
w_\vp\equiv \fr{p_\vp}{\rvp}= \fr{ \xi\dot\vp^2/2 -V }{ \xi\dot\vp^2/2+V}\; ,\hspace{1cm}
w_\phi\equiv \fr{p_\phi}{\rp}= \fr{\le( \vp/q\ri)^{2/3}\xi\dot\phi^2 -M_0^2\phi^2}{ \le( \vp/q\ri)^{2/3}
\xi\dot\phi^2+M_0^2\phi^2}\;.
\ee
Notice that for $\xi=1$, i.e. $\hat\vp<1$, we have $|w_i|\leq 1$, with $ i=\phi,\vp$, however if
$\xi=-1$ ($\hat\vp>1$) we can have $w_\vp<-1$ (or $w_\phi<-1$ ) and therefore these fields could
play the role of phantom fields, which may be favored by
the cosmological data and have been widely studied \ci{phantom}.

In the absence of interaction term and in the
limit of  constant  equation of state
the evolution of the energy densities are given by
$\rho_i\propto a^{-3(1+w_i)}$ however once the interaction term
$\Gamma$ is turned on evolution of $\rho_i$ depends on the effective
equation of state  defined by
\bea
w_{eff \phi}&\equiv&
w_\phi+\fr{\Gamma}{3H\rp}= \fr{9H q^{3/2}\vp^{1/3}[ \le( \vp/q\ri)^{2/3}\xi\dot\phi^2+M_0^2\phi^2]
+ \xi\dot\phi^2\dot\vp}
{9H q^{3/2}\vp^{1/3}(\le( \vp/q\ri)^{2/3}\xi\dot\phi^2+M_0^2\phi^2 )}  \\
w_{eff \vp}&\equiv& w_\vp-\fr{\Gamma}{3H\rvp}=
\fr{9H q^{3/2}\vp^{1/3}[\xi\dot\vp^2/2 -V]-   \xi\dot\phi^2\dot\vp}{9H q^{3/2}\vp^{1/3}(\xi\dot\vp^2/2+V)}
\eea
such that eqs.(\ref{drp}) and (\ref{drvp}) become
\be
\rho_i +3H\rho_i(1+w_{eff i})=0
\ee
with $i=\phi,\vp$.
Which field  will dominate at late times will depend
on which effective equation of state is smaller.

We show in fig.\ref{fig2} an example with $M_0^2= 0.1, q=1$ and $\xi=1$
with initial values $\Om=0.8, \Omvp=0.05, \Omp=0.15$ and final
values $\Om=0.2, \Omvp=0.38, \Omp=0.42$.
The evolution of $\vp$ and $\phi$ (blue, red respectively) oscillate around
the minimum of the potential $\vp_o=q=1$ and $\phi_o=0$. The  energy
densities  $\rp,\rvp$ redshift as matter, i.e. $w=0$. The effect of the
interaction term $\Gamma$ is small and the behavior of $w_\vp, w_\phi$
is similar to  $\wvpe $   and $\wpe$. In this case the dynamics do not
lead to an accelerating epoch of the universe.

We show in fig.\ref{fig3} the same example as before but with   $\xi=-1$.
In this case the fields roll down the potentials $-V$ and $-B$ since
the fields have negative energy kinetic term. The final values
are $\Om=0, \Omvp=0.2 $ and $ \Omp=0.8$.  As in the previous case
the effect of the interaction term $\Gamma$ is small and the behavior of $w_\vp, w_\phi$
is similar to  $\wvpe $   and $\wpe$. However, contrary to the case $\xi=1$, now
we have an eq. of state always smaller than -1. Therefore the universe has
an accelerating behavior at late times once the energy density of $\rvp$ or $\rp$ dominates
the universe. Therefore, $\vp, \phi$ can represent the present day Dark Energy.

\section{Discussion}\label{conclusion}

Starting from the spectrum of the scalar mesons in the ${\cal N}$=2 theory dual to a 
supergravity configuration of D3 and D7 branes, we were able 
to extract the Lagrangian governing the behavior of the two lightest of these states in the 
infrared limit. We considered the possibility for these mesons to live in our universe and 
obtained an action to describe the cosmological consequences of the presence of these fields.

The resulting effective scalar fields $\vp$ and $\phi $ have different cosmological
contributions depending on the sign of the kinetic term. For positive kinetic terms
these fields oscillate around the minimum of the potential but they have different
masses. They  behave as dark matter and therefore their energy densities redshift as
$\rho\propto a^{-3}$. On the other hand, for negative kinetic terms, the  fields
become phantom fields with an equation of state parameter $w<-1$. The universe expands in
an accelerating way and the fields $\vp, \phi$ could then parameterize the Dark Energy.

Concerning the possibility of phantom fields appearing for the case analyzed here,
we have to say we believe it to be a more general phenomenon for this kind of scenarios. The reason
for this believe is that given the Lagrangian in \cite{Hovdebo:2005hm}, we can see that to
each order $n$, higher than two, there will be a contribution given by the product of the kinetic term
of one of the mesons in the spectrum, $\partial_\mu\Phi_a\partial^\mu\Phi_a$, times a polynomial of
homogeneous degree $n-2$ on the other fields. Considering then the Lagrangian to all orders, it would be
found that at least some of the kinetic terms would appear multiplying a polynomial of the fields, and
these fields could take a wide range of values. It would seam natural to expect
then for the kinetic term to be negative in some region of values for the fields, giving rise to
the existence of phantom fields. To make this expectation concrete, it would be necessary to consider
a setting where the physical properties would make the Lagrangian of \cite{Hovdebo:2005hm}, or another one
similarly deduce, exactly summable so that the appearance of phantom fields could be precisely stated.

\section{Acknowledgments}

This work was supported in part by CONACYT project 45178-F  and 2007/4265. LP wants to acknowledge David Mateos 
for the discussion of the present work in its early stage. LP was supported by PROFIP (UNAM, M\'exico).

\end{document}